# Classification of human activity recognition using smartphones

Hoda Sedighi

**Abstract** Smartphones have been the most popular and widely used devices among means of communication. Nowadays, human activity recognition is possible on mobile devices by embedded sensors, which can be exploited to manage user behavior on mobile devices by predicting user activity. To reach this aim, storing activity characteristics, Classification, and mapping them to a learning algorithm was studied in this research. In this study, we applied categorization through deep belief network to test and training data, which resulted in 98.25% correct diagnosis in training data and 93.01% in test data. Therefore, in this study, we prove that the deep belief network is a suitable method for this particular purpose.

**Keywords** Human activity recognition. Deep belief network. Smartphone. Classification.

H.Sedighi

Computer Engineering Department, Faculty of Electrical and Computer Engineering, Shahid Beheshti University G.C., Evin, Tehran, Iran
E-mail: hoda.seddighi@gmail.com

## 1 Introduction

Smartphones and any other mobile devices are becoming an ideal platform for continuous monitoring of user activity because of a large number of sensors embedded in them. Detecting individual activity on smartphones still seems to be a challenge given the limitations of resources such as battery life and computational workload capacity. Considering user activity and managing them, we can conceive low power consumption for mobile phones and other mobile devices, which requires a complete and rigorous program to recognize activities and adjust device power consumption regarding their application at different times and places. However, with the rapid development of new and innovative applications for mobile devices such as smartphones, advances in battery technology do not keep up, especially in energy conservation.

On the other hand, the use of activity recognition is increasing in active and preventive healthcare applications at home, learning environments of security systems, and a variety of human-computer interactions. This paper proposes and implements a system for activity recognition in the home environment with a set of switch sensors and a practical text-based sampling tool. The system uses the extraction and selection techniques to generate the appropriate inputs for the classifier unit. These properties are selected so that they have the highest impact on the recognition and reduce the computational load as much as possible.

Physical activity is any physical movement created by the skeletal muscles that result in energy consumption. Some of these activities include hiking, climbing stairs, exercising, etc. Human physical activities have various facets that can be defined and specified, such as the types of states (such as sitting, standing, walking, lying down), duration and intensity (step speed, power of movement), and frequency (such as the number of mood changes) [1]. People usually do different activities throughout the day. Many of these activities are repeated daily, such as eating around 12 pm, sleeping at about 11 pm, etc. The patterns of daily activity are simply a sequence of activities performed by the user during the day [2].

**2 literature review**

The sensor can collect the data used to recognize human activities. There are three main concerns about the sensor: type, location, and number. In most motion-detector systems, motion sensors, especially accelerometers, are used to estimate body angles from the vertical line and to determine the user's placement and movement. Accelerometers are used to measure linear acceleration (Fig. 1).

The signal obtained from the accelerometer has two components. One is the gravitational acceleration component (static) that is used to collect information about the subject's orientation, and the other is the body (dynamic) acceleration component to gather information about the subject's motion [4]. Three-dimensional accelerometer values are measured in 3 axis units of acceleration in SI (International System of Units) in m/s².

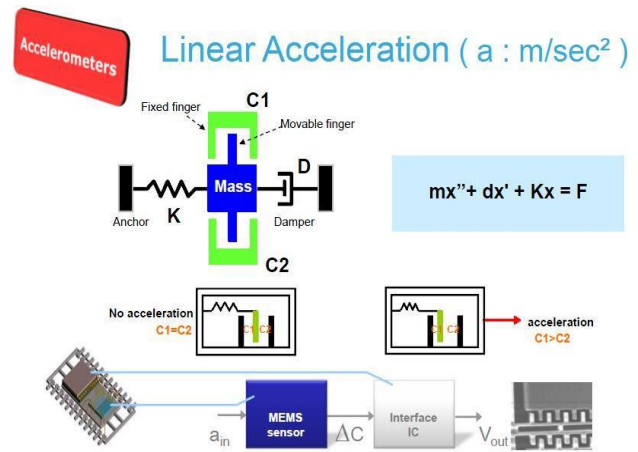

**Fig.1** Structure of linear acceleration sensor[3]

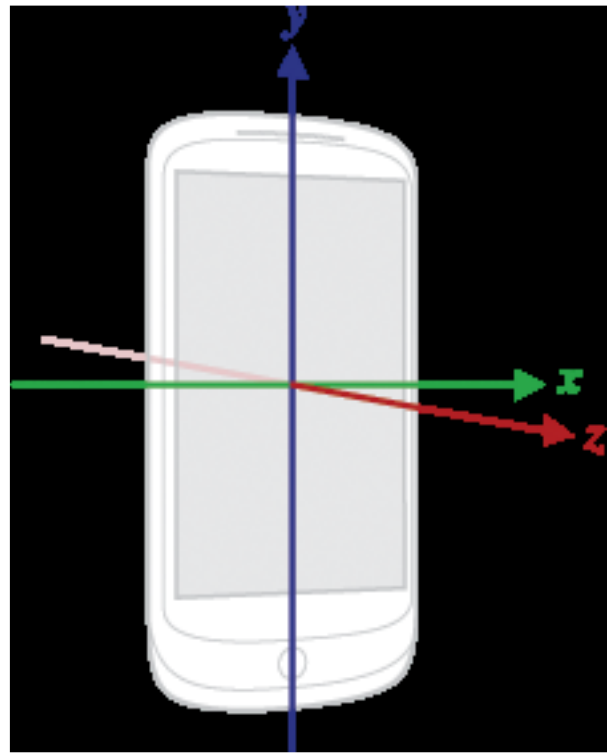

**Fig.2** Accelerometer components [4]

Some researches using the number of activities and computer vision, which are potentially supported by activity recognition and accelerometer, are mentioned in [5-7]. According to [9] researchers are more interested in the use of accelerometers since it is less difficult to collect data from users than other sensors, such as a microphone or camera, it can be used in the background. There are also many other benefits that make this type of sensor useful for studying human activities. For instance, their low cost and small size (which makes it easy to embed them in the phones).

Previous researches have shown that movements such as walking, running, and climbing stairs, and states such as sitting, standing and sleeping, can be illustrated 83% to 95% accurately using hip, thigh, and wrist acceleration. However, previous research [10] has shown that thighs and wrists are often better places for accelerometers to detect activities. Often the wrist accelerometer data is useful for determining the movements of the upper body, and the accelerometer data on the thigh helps discriminate the activities of the limbs. Activity recognition systems that work with data from different locations allow the user to transport the device to the right place to work in the field.

[11] Also presents a system called TAHAR (Transition-Aware Human Activity Recognition), in which the system architecture is used to recognize physical activities using smartphones. It targets real-time classification with a set of motionless sensors while addressing the problem of transitions between known and unknown activities to the learning algorithm. This is accomplished by combining the probable return performance prediction of a support vector machine (SVM) with a heuristic filtering method.

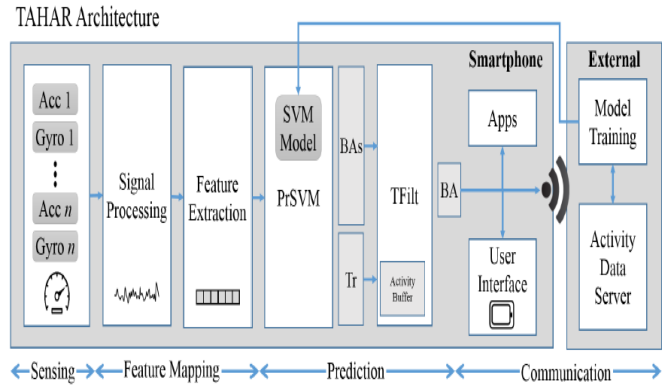

**Fig.3** TAHAR system [11]

**3 Deep belief networks**

Belief networks are made up of some layers of restricted Boltzmann machines. The restricted Boltzmann machine is also a kind of Boltzmann machine which consists of the connections between some hidden units and some obvious ones. The Boltzmann machine is a kind of unidirectional graphical model, which is also called the Markov random field or MRF [12]. At the heart of deep belief network is a unidirectional graphical model that is termed the restricted Boltzmann machine.

The idea of the greedy learning algorithm for deep belief networks is relatively simple. The learning algorithm uses a restricted Boltzmann machine set, as shown in Figure 4. First, it trains the lower restricted Boltzmann machine using parameters of $\boldsymbol{W}^1$. Then, the second layer weights are initialized with $\boldsymbol{W}^2=\boldsymbol{W}^{1T}$ to ensure that both hidden layers of the deep belief network are at least as good as the original restricted Boltzmann machine. After extracting the first hidden layer ($h^1$) values, the deep belief network can be improved by using this data and $\boldsymbol{W}^2$ correction. [13] However, in general state, the size of the weighing matrix in $\boldsymbol{W}^1$ and $\boldsymbol{W}^2$ does not need to be the same.

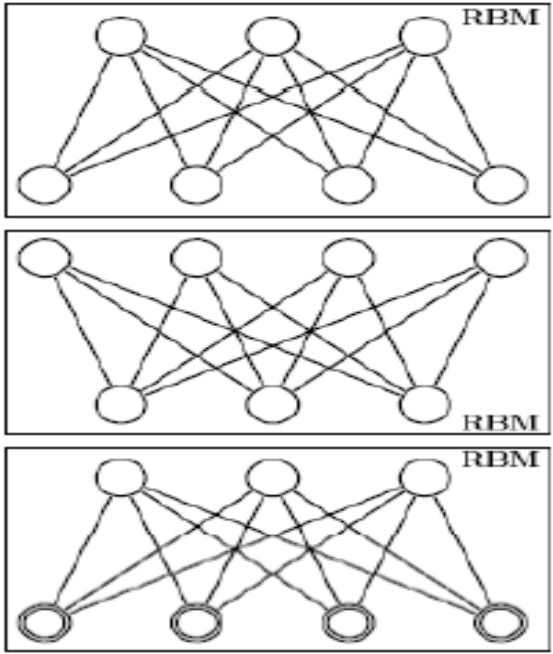

**Fig.4** Greedy learning of stacks from restricted Boltzmann Machines where Lower Level Samples of restricted Boltzmann Machines are used as Data for the Next restricted Boltzmann Machine's Training [13]

This idea can also be used to train the third layer of the Boltzmann restricted machine on the $h^2$ vector, which is derived from the second restricted Boltzmann machine. By setting $W^3 = W^{2T}$ it is guaranteed that the lower bound of the likelihood logarithm improves. We can also improve boundaries by changing $W^3$. This layer-by-layer method can be repeated several times to obtain a deep hierarchical model. This method is seen in the algorithm of Figure 5.

### 3.1 Restricted Boltzmann Machine

Restricted Boltzmann machines are a special kind of Markov random fields. A Boltzmann machine is a symmetric network of binary random units. This network has a set of visible

1. Assigning $w^1$ parameters from the first layer of RBM to the data

2. Fixing $w^1$ parameter and using $h^1$ samples from $Q(h^1|v) = P(h^1|v;W^1)$ as data for teaching binary features of next layer in an RBM

3. Fixing $w^2$ which defines the second layer of teachers and using samples of $h^2$ from $Q(h^2|h^1) = P(h^2|h^1;W^2)$ as data for teaching binary features of the third layer

4. Continuing this process recursively for next layers

**Fig.5** Recursive greedy learning process in DBN [13]

units of $v \in \{0,1\}^{g_v}$ and a set of hidden units of $h \in \{0,1\}^{g_h}$, in which are $g_v$ and $g_h$ are the number of visible units and hidden units, respectively. The energy of $\{v, h\}$ state in the Boltzmann machine is defined as Eq. 1:

$$E(v,h) = -\frac{1}{2}v^T L v - \frac{1}{2}h^T J h - v^T W h \quad (1)$$

Due to the complexity of the Boltzmann machine calculations, a simpler type, which is called the restricted Boltzmann machine, would be implemented. Setting $J=0$ and $L=0$ result in the famous restricted Boltzmann machine model. (i.e., the image on the right side of Figure 6)

In a restricted Boltzmann machine the energy of $\{v, h\}$ state, considering the bias, is equal to:

$$E(v,h) = -v^T W h - a^T v - b^T h$$
$$= -\sum_{i=1}^{g_v}\sum_{j=1}^{g_h} W_{ij} v_i h_j - \sum_{i=1}^{g_v} a_i v_i - \sum_{j=1}^{g_h} b_j h_j \quad (2)$$

In which $W_{ij}$ represents the symmetric transactional parameter between the observable unit of $i$ and hidden units of $j$, and $b_i$ and $a_j$

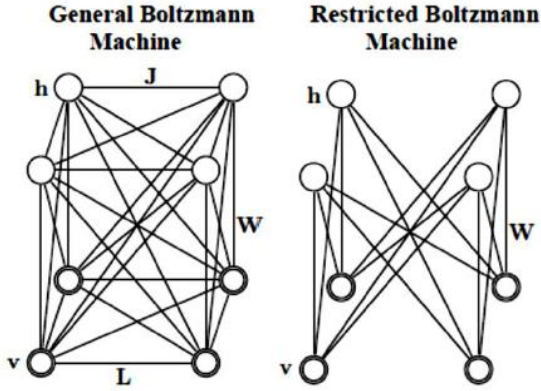

**Fig.6** on the Left, A general Boltzmann machine is shown. The top layer represents random binary hidden properties, and the bottom layer represents vectors of random binary variables. On the right, a restricted Boltzmann machine, with no connections between hidden and visible units, is illustrated. [14]

represent the bias parameter for the hidden and visible units, respectively. This network assigns a probability value to any possible values of the visible and hidden vectors with the energy function. The definition of potential functions in MRFs and their positive characteristic lead us to use the potential function exponentially in the form of Boltzmann distribution. The joint-probability distribution is defined by multiplying the potential functions, and in this way, the total energy would be equal to the sum of the energies of the potential functions. Accordingly, the joint probability distribution on visible and hidden units is defined as Eq. 3:

$$P(v,h) = \frac{1}{Z}\exp(-E(v,h)) \quad (3)$$

The value of $Z$ is known as the normalization constant or division function, which is obtained from the sum of all visible and hidden vector states.

$$Z = \sum_v \sum_h \exp(-E(v,h)) \quad (4)$$

The value of $Z$ is known as the normalization constant or division function. The probability that the model assigns to the visible v vector equals to the sum of all possible hidden vectors of $h$.

$$P(v) = \sum_h P(v,h) = \frac{1}{Z}\sum_h \exp(-E(v,h)) \quad (5)$$

The probability that the network attributes to the training data can be increased by adjusting the weights and bias to achieve less energy for this data and more energy for other data, which causes changes in $Z$. Accordingly, and to find the appropriate parameters, the objective function is defined as Eq. 6:

$$\text{maximize}_{\{w_{ij},a_i,b_j\}} \frac{1}{m}\sum_{l=1}^{m} \log\left(\sum_h P(v^{(l)}, h^{(l)})\right) \quad (6)$$

Which $m$ represents the number of training samples, and the aim is to increase the probability of this model for training data. For this purpose, by calculating the partial derivative of this statement with $wij$, we will have:

$$\frac{\partial}{\partial w_{ij}}\left(\frac{1}{m}\sum_{l=1}^{m}\log\left(\sum_h P(v^{(l)}, h^{(l)})\right)\right) \quad (7)$$

$$= \frac{1}{m}\sum_{l=1}^{m}\sum_h X_{il}h_j P(h|v=x) - \sum_{v'}\sum_{h'} v'_i h'_j P(v',h')$$

Where $X_{il}$ represents the $i^{th}$ unit (or dimension) of the $l^{th}$ training data. The first expression in the above derivative can be calculated precisely but calculating the second expression is not possible. Therefore, other methods of calculation would be considered for this purpose. Logarithm derivative of the probability for training data to the network weights can be calculated with an estimation as follows.

$$\frac{\partial \log P(v)}{\partial w_{ij}} = <v_i h_j>_{data} - <v_i h_j>_{model} \quad (8)$$

**4 Evaluation**

In this phase of the study, we used SBHAR for the data set. In this dataset, the number of input data for training equals 7767 samples (each sample/record containing 12 data). For the test data, also, we have 3162 data. The total number of output classes is 12. These 12 activity classes are as follows.

**Table 1:** 12 classes of activities in dataset

| # | Activity | # | Activity |
|---|---|---|---|
| 1 | WALKING | 7 | STAND_TO_SIT |
| 2 | WALKING_UPSTAIRS | 8 | SIT_TO_STAND |
| 3 | WALKING_DOWNSTAIRS | 9 | SIT_TO_LIE |
| 4 | SITTING | 10 | LIE_TO_SIT |
| 5 | STANDING | 11 | STAND_TO_LIE |
| 6 | LAYING | 12 | LIE_TO_STAND |

The data were obtained from a group of 30 participants aged 19-48 years. To do the experiment, we used 50 neurons in the first and second hidden layers and 10 neurons in the third layer. We also performed 25 generations of processes for each layer.

Figure 7 shows the first implementation of Deep Belief Network. The results we achieved were significant and impressive. In the first run, we achieved 2.16% of error and recognition of 97.84%. This correct recognition is for precision in training.

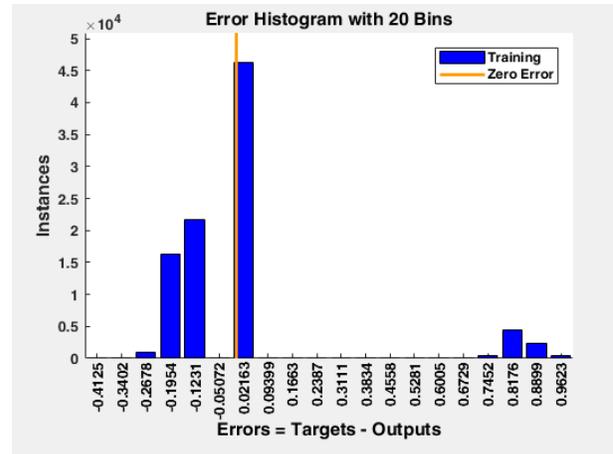

**Fig.8** Precision of Training in the First implementation

On the test data, we also achieved 6.99% error and 93.01% correct recognition.

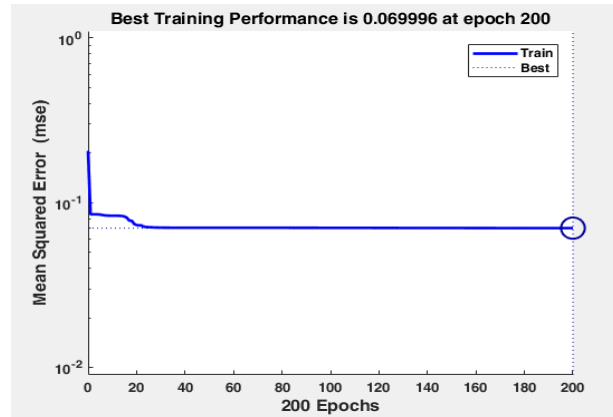

**Fig.9** precision rate at first implementation on test data

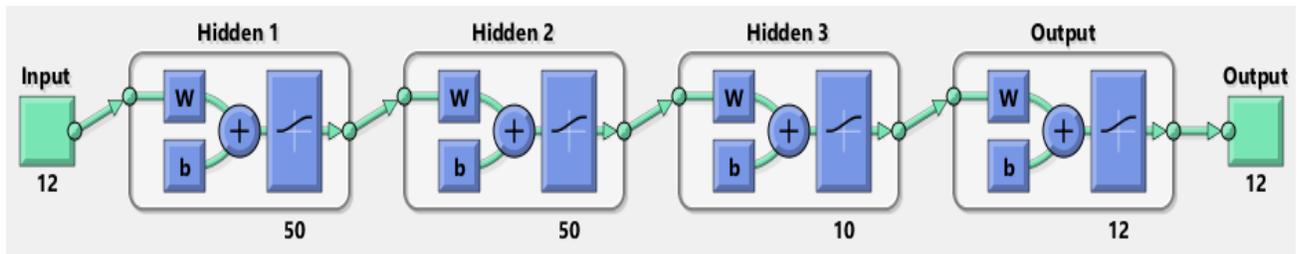

**Fig.7** the First Implementation of Deep Belief Network

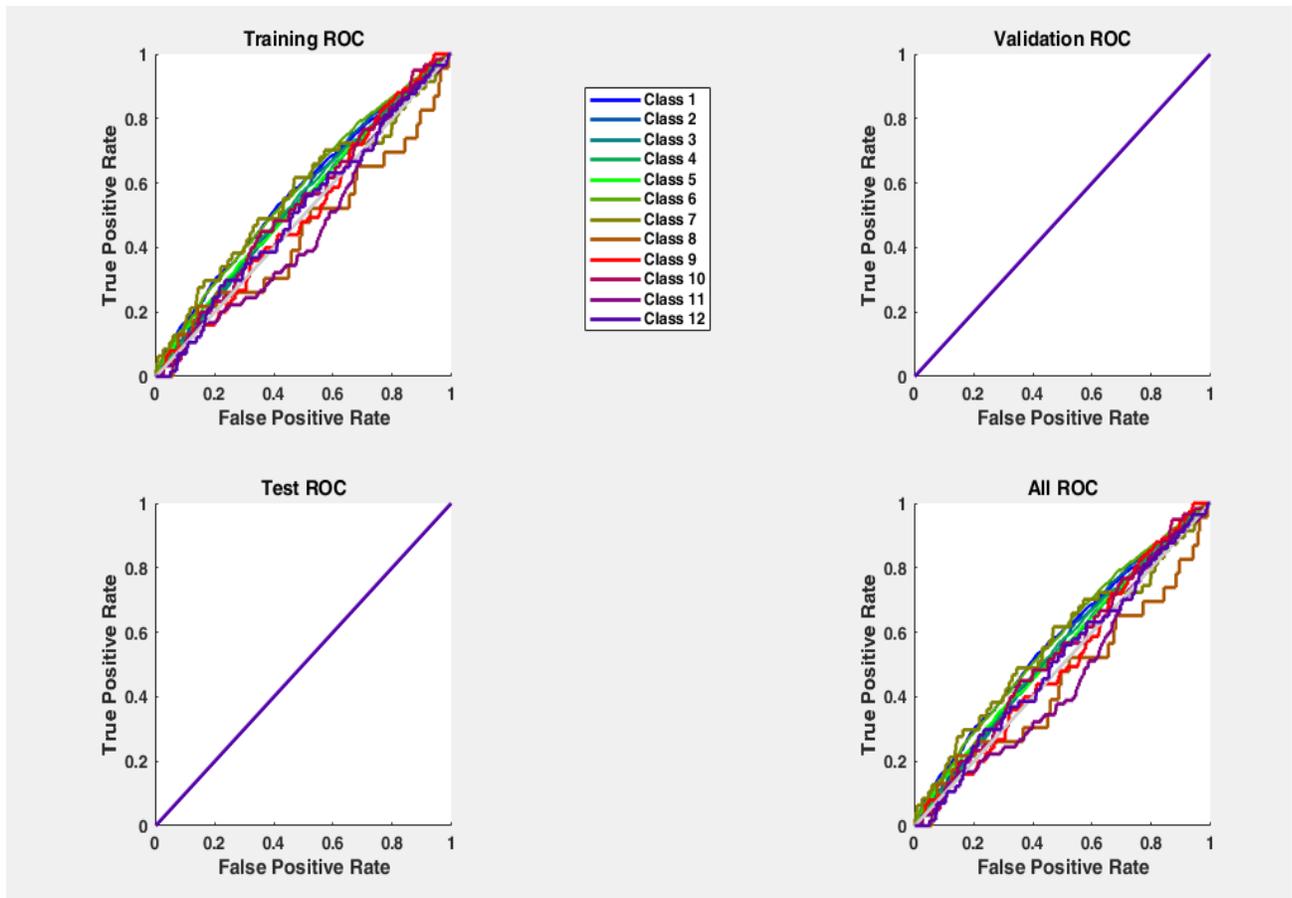

**Fig.10** ROC diagram at first implementation

This precision is also evident in the ROC chart. For the second time, we considered 1000 neurons per layer and also processed up to 250 generations per layer. The first time, results were obtained rather quickly within 3 seconds. However, For the Second time, calculations took several minutes (over 25 minutes).

We achieved significant and impressive results. In the first run, we achieved 1.75% error and a recognition of 98.25%. This is the correct recognition for the precision of training.

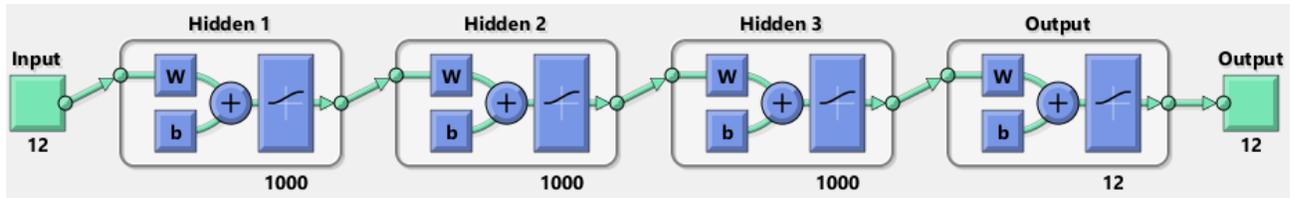

**Fig.11** Second Implementation of Deep Belief Network

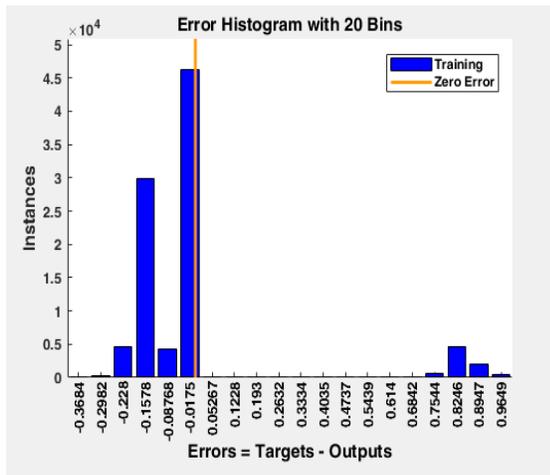

**Fig.12** Precision of Training in Second implementation

On the test data, we also achieved 7.01% error and 92.99% correct recognition, which is only 0.02 % less than the recognition and precision of the first implementation.

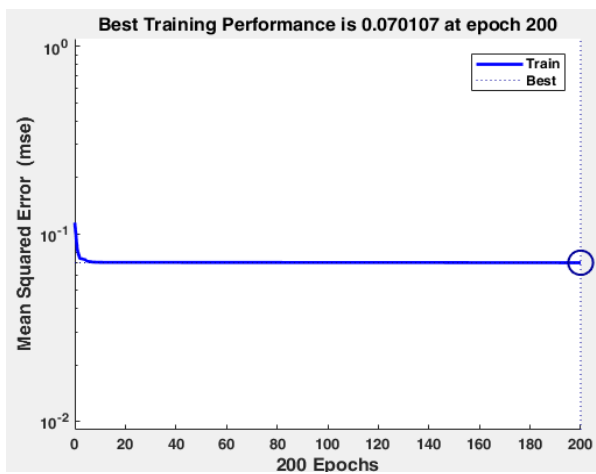

**Fig.13** precision of the second implementation on the test data

## 5 Conclusion

Smartphones and any other mobile devices are becoming an ideal platform for continuous monitoring of users' activities using a large number of sensors embedded in them. Recognizing individual activities on smartphones still seems to be farfetched, given the limited resources such as battery life and computational workloads. By Considering users' activities and managing them, low power consumption can be achieved for smartphones and other mobile devices, however; it requires a complete and meticulous planning to recognize activities and adjust energy consumption regarding their functions, time and place of usage. But with the rapid improvements of new and novel applications of mobile devices such as smartphones, advances in battery technology and Energy storage do not keep up. This is just one of the applications for monitoring and recognizing individual activity on smartphones. In this study, we tried to categorize activities using a deep belief network; the results are very promising, and in comparison to the previous studies, remarkable progress can be noticed. Comparison of this study with similar studies presents in table 2.

According to the table 2, the results of the current study are well comparable with similar cases both in terms of running time (for implementing with the low number of neurons) and in the precision of recognition.

**Table 2:** Comparison of this study with similar studies

| # | Classification method | Accuracy of detection | # | Classification method | Accuracy of detection |
|---|---|---|---|---|---|
| 1 | [16] DTW + KNN | 69.49 | 7 | [19] Actionlet + LOP + Pyramid | 82.2 |
| 2 | [16] LSTM | 85.82 | 8 | [16] BoF + SVM | 85.56 |
| 3 | [16] dRNN | 87.1 | 9 | [16] TemporalAggregate | 89.27 |
| 4 | [18] Depth Motion Maps | 66.11 | 10 | [17] FTA + KNN | 90.69 |
| 5 | [16] pFTA-Random + KNN | 88.7 | 11 | Proposed method for training data | 98.25 |
| 6 | [16] pFTA-Learn + KNN | 90.96 | 12 | Proposed method for test data | 93.01 |